\documentclass[11pt,a4paper]{article}

\usepackage[T1]{fontenc}
\usepackage[utf8]{inputenc}
\usepackage{lmodern}
\usepackage[margin=2.5cm]{geometry}
\usepackage{amsmath,amssymb}
\usepackage{graphicx}
\usepackage{booktabs,array}
\usepackage{authblk}
\usepackage[authoryear,round]{natbib}
\usepackage[hidelinks]{hyperref}

\graphicspath{{./}{figures/}}
\title{Event-Aware Spatiotemporal Precipitation Forecasting with Geographic Context and Physics-Guided Regularization}
\author[1,2]{Yiping Hong}
\author[1]{Xinyu Wang}
\author[3]{Sameh Abdulah\thanks{Corresponding author. Email: \href{mailto:sameh.abdulah@kaust.edu.sa}{\texttt{sameh.abdulah@kaust.edu.sa}}}}
\affil[1]{School of Mathematics and Statistics, Beijing Institute of Technology, Beijing, China}
\affil[2]{Intelligent Innovation and Interdisciplinary Integration Base, Tangshan Research Institute, Beijing Institute of Technology, Hebei, China}
\affil[3]{Computer, Electrical and Mathematical Sciences and Engineering (CEMSE) Division, King Abdullah University of Science and Technology, Thuwal, Saudi Arabia}
\date{}

\begin{document}
\maketitle

\begin{abstract}
Hourly precipitation forecasting involves several distinct statistical challenges, including spatially varying predictor–precipitation relationships, a strongly imbalanced precipitation distribution, and progressive degradation of precipitation event skill with increasing lead time. 
We develop a spatiotemporal forecasting framework that addresses these challenges through three complementary components: explicit geographic representation, event-aware learning for imbalanced precipitation, and weak asymmetric regularization derived from the atmospheric water budget.
The physical information is treated as an asymmetric constraint rather than as an additional prediction target, designed to discourage physically unsupported precipitation attenuation without replacing the data-driven forecast. 
Experiments using ERA5 data across regional, enlarged domain, and spatial subset settings show that explicit geographic information improves spatial field prediction, while event-aware learning provides the most consistent gains in detecting moderate and heavy precipitation events. 
Physical regularization has a more selective effect, mainly reducing systematic underprediction over the enlarged domain while improving longer lead precipitation event prediction in the spatial subset experiment.
These results indicate that the benefit of physical guidance depends on the available data regime and becomes most apparent when data-driven precipitation information deteriorates with increasing lead time.
\end{abstract}


\section{Introduction}

Spatial prediction infers unobserved values from incomplete measurements by exploiting spatial dependence and relevant auxiliary information. Among classical geostatistical approaches, kriging leverages the spatial correlation among observations to predict values at unobserved locations, while regression kriging extends this framework by combining covariate driven modeling of large scale spatial variation with kriging of the remaining spatially correlated residuals \citep{hengl2007regression,mohammadpour2024machine}. This combination is particularly relevant for precipitation prediction, where environmental and geographical covariates can capture broad spatial patterns while neighboring observations provide complementary information about local variability \citep{goovaerts2000geostatistical,dura2024spatial}.

Environmental and geophysical processes, however, often exhibit spatial behavior that is inconsistent with stationary modeling assumptions. Their variability and dependence structures can change substantially across the study domain, motivating nonstationary covariance models based on spatially varying parameters and localized representations \citep{paciorek2006spatial,chang2010semiparametric}. This spatial nonstationarity becomes even more pronounced in multivariate climate settings, where not only the marginal dependence of individual variables but also their cross dependence can vary substantially across geographic regions \citep{jun2014matern}.

Beyond these spatially varying dependence structures, certain climate and weather variables, such as precipitation, introduce additional distributional challenges that further complicate prediction. Rainfall observations typically contain a large fraction of zero or near-zero values, while positive amounts are strongly oriented to the right, motivating models that explicitly distinguish precipitation occurrence from precipitation intensity \citep{berrocal2008probabilistic}. At hourly temporal resolution, extreme precipitation can exhibit particularly heavy tailed behavior together with coherent spatial structure \citep{https://doi.org/10.1029/2018WR022732}. Consequently, short term precipitation forecasting must address several challenges simultaneously, including spatially heterogeneous dependence, severe imbalance between dry and wet conditions, heavy tailed precipitation intensities, and progressively weaker predictive information as the forecast horizon increases.

While classical geostatistical models provide explicit mechanisms for representing spatial dependence and nonstationarity, modern precipitation forecasting extends beyond spatial interpolation at a single time point \citep{bi2023pangu}. The target is a dynamically evolving spatial field whose future states depend jointly on recent precipitation and multiple atmospheric covariates distributed over space and time. This leads to a problem of high dimensional nonlinear spatiotemporal regression, and specifying and estimating a complete covariance structure becomes increasingly difficult as the spatial and temporal dimensions grow \citep{wang2025spatialcnn}. Similar challenges arise in large environmental datasets, where spatial heterogeneity and high dimensionality impose substantial modeling and computational burdens on conventional Gaussian process approaches \citep{salvana2022parallel,konomi2014adaptive,pan2024gpu}.

Deep learning offers an alternative way to learn nonlinear dependence directly from large gridded datasets \citep{LI2023100777}. Recent data-driven weather systems trained on reanalysis data show the scalability of these approaches for high dimensional atmospheric forecasting \citep{bi2023pangu,rasp2024weatherbench2}, while convolutional recurrent architectures such as ConvLSTM frame precipitation forecasting as a spatiotemporal sequence prediction problem \citep{shi2015convlstm}. However, replacing explicit covariance modeling with neural representations does not remove the underlying statistical structure of the problem. Hourly precipitation forecasting still requires a model to represent spatially varying relationships, accommodate a strongly imbalanced response distribution, and incorporate imperfect physical information without treating it as exact observational truth. We therefore view neural precipitation forecasting as a structured spatiotemporal prediction problem rather than solely as a flexible function approximation task. Section~2 develops these statistical and physical considerations in more detail.

Taken together, existing studies have made substantial progress in neural spatial representation, event-oriented precipitation learning, and physics informed modeling, but these sources of information are often considered separately. We therefore develop a two--stage spatiotemporal forecasting framework that integrates geographic context, precipitation occurrence--intensity structure, and uncertain physical information. The first stage learns the data-driven precipitation field from atmospheric covariates, explicit geographic information, and an event-oriented objective. The second stage introduces the atmospheric water budget diagnostic as weak asymmetric evidence through structured one sided regularization. This separation allows physical information to discourage attenuation of unsupported precipitation without forcing the forecast to reproduce the diagnosis pointwise.

This study makes three main contributions. First, we distinguish explicit geographic context from local spatiotemporal dependence by incorporating absolute location information alongside convolutional recurrent features. Second, we develop an occurrence-intensity-aware learning objective that combines robust intensity estimation, differentiable event-oriented learning, and auxiliary occurrence prediction to address the strongly imbalanced precipitation distribution. Third, we introduce an uncertainty aware physics-guided fine tuning strategy in which atmospheric water budget information is treated as asymmetric physical evidence rather than as a deterministic precipitation target, with the physical guidance allowed to adapt to different forecast stages and data regimes. We evaluate the framework across multiple spatial domain and forecast horizon settings, including the focal loss configuration of GENESIS ($\mathrm{GENESIS}_{\mathrm{FL}}$) as an event-oriented ConvLSTM benchmark \citep{rahimi2024genesis}. The experiments assess both predictive performance and when physical guidance adds value.

The remainder of this paper is organized as follows. Section~2 discusses existing statistical and physical challenges that motivate the proposed framework, including spatial heterogeneity, precipitation imbalance, and the use of physical guidance at longer forecast horizons. Section~3 introduces the spatiotemporal problem formulation, geographic representation, recursive forecasting architecture, occurrence-intensity-aware learning objective, and physics-guided regularization. Section~4 describes the data, experimental configurations, comparison models, and evaluation metrics. Section~5 presents the forecasting results across spatial domains and forecast horizons. Section~6 discusses the statistical and physical implications of the findings, and Section~7 summarizes the main conclusions and future directions.

\section{Statistical and Physical Motivation}
Short-term precipitation forecasting involves several interrelated statistical and physical challenges that standard spatiotemporal learning objectives do not explicitly address. In particular, accurate prediction requires accounting for spatial heterogeneity, the highly imbalanced distribution of precipitation, and the progressive loss of predictive information at longer forecast horizons. These challenges motivate the modeling choices developed in the following subsections.

\subsection{Spatial Dependence and Geographic Heterogeneity}

One challenge lies in distinguishing between spatial dependence and spatial heterogeneity. Spatial dependence describes association among observations at different locations, whereas spatial heterogeneity allows the relationship between predictors and the response to change across the spatial domain. 
The latter has long motivated nonstationary and geographically weighted regression, and recent neural extensions have used nonlinear spatial or spatiotemporal weighting structures to represent more complex location-dependent relationships \citep{wu2021gtnnwr}. Related work has also demonstrated how convolutional neural networks can efficiently estimate spatially varying covariance structures in large scale nonstationary spatial models \citep{nag2025efficient}. Statistical deep learning models have similarly begun to incorporate geographic information explicitly through direct spatial inputs, learned spatial embeddings, and spatially varying model components \citep{jahn2024flexible,zammit2024spatial,zhan2025neural}.

Related studies on neural spatial prediction have shown that learned geographic representations can be combined with ordinary attribute features to provide location-specific information not captured by neighborhood structure alone \citep{bao2024twostage,ma2025geographically}. These developments suggest that learning local spatial patterns and representing absolute geographic context need not be the same modeling task. In regularly gridded atmospheric fields, shared convolutional operators efficiently extract recurring local structures, but the convolution operation itself does not explicitly identify where those structures occur. This motivates supplementing convolution-based local dependence learning with a lightweight representation of geographic location, thereby allowing otherwise similar atmospheric configurations to be distinguished by their spatial context.

\subsection{Imbalanced Precipitation and Event-Oriented Learning}

A second challenge arises from the strongly imbalanced distribution of hourly precipitation. Standard regression objectives aggregate errors across the empirical distribution and are therefore dominated by the large number of zero, near-zero, and weak precipitation samples. Such optimization can favor predictions close to frequently observed values while attenuating relatively rare high-intensity events. More generally, deterministic data-driven weather models have been observed to generate smoother spatial fields, with a corresponding risk of underestimating the magnitude of extreme events \citep{bi2023pangu}. For precipitation, this implies that favorable overall regression accuracy does not necessarily correspond to strong threshold-based event skill. The GENESIS model provides a relevant example: its U-Net--ConvLSTM architecture was evaluated with both mean squared-error regression and focal loss classification, with the classification configuration showing stronger skill for high-intensity precipitation while the regression model performed better for light rainfall \citep{rahimi2024genesis}. The same study also showed that including physically relevant variables, such as precipitable water and horizontal winds, becomes increasingly useful for predicting intense precipitation at longer lead times. Related physics-constrained precipitation forecasting has similarly reported that categorical objectives can complement regression when longer lead forecasts underestimate precipitation intensity \citep{wang2026physics}. These results indicate that precipitation occurrence, continuous intensity, and threshold-based event skill capture complementary aspects of forecast performance. A learning objective designed for hourly precipitation should therefore account for these components jointly rather than relying on a single average error criterion.

\subsection{Physical Guidance for Longer-Lead Forecasting}

A third challenge concerns the progressive degradation of forecast skill with increasing lead time and whether physical information can mitigate it. Precipitation development is closely related to atmospheric moisture availability and transport, which operate across multiple spatial and temporal scales. Analyses over East Asia have demonstrated strong relationships between precipitation variability and atmospheric moisture transport, providing a physical basis for using the atmospheric moisture budget to characterize precipitation-supporting conditions \citep{kukulies2024moisture}. More broadly, researchers can incorporate physical knowledge into statistical models through regularization rather than treating it as an additional observation. For example, \citet{desanctis2026physics} embedded an advection-diffusion partial differential equation into semiparametric spatiotemporal regression, balancing data fidelity against physically driven dynamics. Physics-guided deep learning models have likewise incorporated moisture transport into precipitation forecasting; for example, PS2F-Net constrains vertically integrated water vapor transport and reports improved precipitation event skill and forecast stability at extended lead times \citep{wang2026physics}. These studies demonstrate the potential value of combining data-driven spatiotemporal learning with physically meaningful process information.

For short-term precipitation forecasting, however, a moisture budget diagnostic should not necessarily be interpreted as an exact precipitation target. Temporal and spatial discretization, uncertainty in reanalysis variables, unresolved atmospheric processes, and scale mismatch can produce appreciable discrepancies between budget derived and observed precipitation. Directly forcing neural predictions toward such a diagnostic may therefore transfer physical diagnostic errors into the forecast. The relevant question is instead how imperfect physical information can constrain a data-driven prediction without being treated as deterministic truth. In particular, when the moisture budget strongly supports precipitation but the model predicts little or no rainfall, the forecast may be inconsistent with the available physical evidence. The converse is less definitive: the budget diagnostic may underestimate or spatially misplace precipitation and therefore does not provide a reliable upper bound on the forecast. This asymmetry motivates incorporating the physical diagnostic through a one-sided penalty rather than enforcing symmetric agreement. The forecast is penalized when it falls substantially below the precipitation level supported by the calibrated diagnostic, but it is not forced to reproduce the diagnostic magnitude or location exactly. Limited spatial displacement and temporal redistribution can therefore be accommodated while discouraging physically unsupported attenuation at longer lead times.

These three challenges directly inform the design of the proposed forecasting framework. Spatial heterogeneity motivates explicitly incorporating geographic information alongside convolutional spatiotemporal features; precipitation imbalance motivates an occurrence-intensity-aware learning objective; and forecast skill degradation at longer lead times motivates using atmospheric water budget information as a weak, asymmetric physical regularizer. Building on these considerations, we next formulate the forecasting problem and describe the methodological components introduced to address each challenge.

\section{Methodology}
We now present the proposed spatiotemporal forecasting framework and its main methodological components. These include the problem formulation, geographic representation, latent spatiotemporal modeling, event-aware learning, and physics-guided fine-tuning.

\subsection{Spatiotemporal problem formulation}
\label{sec:problem_formulation}

Following the statistical and physical considerations outlined above, we formulate hourly precipitation forecasting as a nonnegative spatiotemporal prediction problem and describe the geographic, event-aware, and physics-guided components of the proposed framework. Let
\begin{equation}
\mathcal{S}=\{\mathbf{s}_1,\ldots,\mathbf{s}_N\},
\qquad
\mathbf{s}_i=(\phi_i,\lambda_i).
\end{equation}
denote the spatial grid, where $\phi_i$ and $\lambda_i$ represent the latitude and longitude of location $\mathbf{s}_i$, respectively. Let $Y_t(\mathbf{s})$ denote hourly precipitation at location $\mathbf{s}$ and time $t$. The dynamic atmospheric state is written in the general form
\begin{equation}
\mathbf{X}_t(\mathbf{s})=
\left[Y_t(\mathbf{s}),A_{1,t}(\mathbf{s}),\ldots,A_{P,t}(\mathbf{s})\right]^{\mathrm{T}},
\label{eq:dynamic_predictors}
\end{equation}
where $A_{p,t}$ denotes the $p$th atmospheric predictor. This notation accommodates the predictor sets used in different experiments, while the experimental design section specifies the exact variables. Given $L$ historical fields,
\begin{equation}
\mathcal{H}_t=(\mathbf{X}_{t-L+1},\ldots,\mathbf{X}_{t}),
\label{eq:historical_sequence}
\end{equation}
Given the historical sequence and static geographic information, the objective is to learn a nonlinear mapping that generates the next $K$ precipitation fields. For forecast horizon $k$,
\begin{equation}
\widehat{Y}_{t+k}(\mathbf{s})
=
f_k\!\left(
\mathcal{H}_t,
\mathbf{g}(\mathbf{s});
\boldsymbol{\theta}
\right),
\qquad k=1,\ldots,K,
\label{eq:predictive_mapping}
\end{equation}
where $f_k(\cdot;\boldsymbol{\theta})$ denotes the learned nonlinear predictive mapping for the $k$th forecast horizon, $\boldsymbol{\theta}$ represents the model parameters, and $\mathbf{g}(\mathbf{s})$ (defined in Section 3.2) denotes static geographic information. The values of $L$ and $K$ are specified according to the experimental configuration described in Section~4.

Because hourly precipitation is strongly right skewed, precipitation is first transformed using

\begin{equation}
Y_t^{\log}(\mathbf{s})
=
\log\left[1+Y_t(\mathbf{s})\right].
\label{eq:log_precipitation}
\end{equation}

The log transformed precipitation is then standardized as

\begin{equation}
Y_t^{*}(\mathbf{s})
=
\frac{
Y_t^{\log}(\mathbf{s})-\mu_Y
}{
\sigma_Y
},
\label{eq:standardized_precipitation}
\end{equation}
where $\mu_Y$ and $\sigma_Y$ are the mean and standard deviation of the log transformed precipitation computed from the training set. The remaining predictors are standardized using their corresponding training set statistics, which are retained for validation and testing.

\subsection{Spatially explicit geographic representation}
\label{sec:geo_representation}

Shared convolutional filters efficiently represent recurring local patterns but do not explicitly identify where those patterns occur. To allow the forecasting relationship to vary with geographic location, six static coordinate features are constructed for each grid cell:
\begin{equation}
\mathbf{g}(\mathbf{s}_i)=
\left[
\widetilde{\phi}_i,
\widetilde{\lambda}_i,
\sin\left(\frac{\pi\phi_i}{180}\right),
\cos\left(\frac{\pi\phi_i}{180}\right),
\sin\left(\frac{\pi\lambda_i}{180}\right),
\cos\left(\frac{\pi\lambda_i}{180}\right)
\right]^{\mathrm{T}},
\label{eq:geo_features}
\end{equation}
where $\widetilde{\phi}$ and $\widetilde{\lambda}$ are standardized latitude and longitude.

A pointwise geographic embedding network maps these six features to an $d_{\mathrm{geo}}$ channel representation, where $\theta_{\mathrm{geo}}$ denotes the trainable parameters of the geographic embedding network and $d_{\mathrm{geo}}$ denotes the number of geographic embedding channels and is set to 8 in this study.
\begin{equation}
\mathbf{e}(\mathbf{s})=f_{\mathrm{geo}}\left(\mathbf{g}(\mathbf{s});\theta_{\mathrm{geo}}\right),
\label{eq:geo_embedding}
\end{equation}
implemented by two $1\times1$ convolutional layers with dimensions $6\rightarrow16\rightarrow d_{\mathrm{geo}}$ and ReLU activations. The embedding is shared across time and concatenated with the dynamic atmospheric field,
\begin{equation}
\widetilde{\mathbf{X}}_t(\mathbf{s})=
\left[\mathbf{X}_t(\mathbf{s}),\mathbf{e}(\mathbf{s})\right].
\label{eq:geo_concat}
\end{equation}
Thus, the geographic embedding represents absolute location, while the spatial encoder learns local interactions between neighboring grid cells.

\subsection{Latent spatiotemporal representation}
\label{sec:latent_representation}

For each historical time step, a shared CNN encoder maps the geographically augmented field to a latent spatial representation,
\begin{equation}
\mathbf{Z}_j=f_{\mathrm{enc}}\left(\widetilde{\mathbf{X}}_j;\theta_{\mathrm{enc}}\right),
\qquad j=t-L+1,\ldots,t.
\label{eq:cnn_encoder}
\end{equation}
The encoder consists of three $3\times3$ convolutions with channel dimensions $(C+d_{\mathrm{geo}})\rightarrow32\rightarrow64\rightarrow64$. Here, $d_{\mathrm{geo}}$ denotes the number of geographic embedding channels introduced in Eq.~(\ref{eq:geo_embedding}). The geographic embedding is concatenated with the $C$ dynamic atmospheric variables before convolution. In this study, $d_{\mathrm{geo}}=8$. The second layer uses stride two, reducing the spatial resolution from $H\times W$ to $H/2\times W/2$. The resulting latent field therefore has 64 channels and preserves local spatial structure at reduced resolution.

The latent sequence $\mathbf{Z}_{t-L+1},\ldots,\mathbf{Z}_t$ is then processed by a ConvLSTM,
\begin{equation}
(\mathbf{H}_j,\mathbf{C}_j)
=F_{\mathrm{CLSTM}}\left(\mathbf{Z}_j,\mathbf{H}_{j-1},\mathbf{C}_{j-1}\right),
\label{eq:convlstm_compact}
\end{equation}
where $\mathbf{H}_j$ and $\mathbf{C}_j$ denote the hidden and cell states. The parameter $\theta_{\mathrm{enc}}$ represents the trainable parameters of the shared CNN encoder. The ConvLSTM uses $3\times3$ convolutions and 64 hidden channels, so the recurrent states remain spatial fields rather than vectors. The terminal states $(\mathbf{H}_t,\mathbf{C}_t)$ summarize the historical spatiotemporal context and initialize future forecasting.

This design separates two complementary sources of spatial information: the geographic embedding represents absolute location and spatial heterogeneity, whereas the CNN--ConvLSTM backbone represents local spatial dependence and its temporal evolution.

\subsection{Recursive multi step precipitation forecasting}
\label{sec:recursive_forecasting}

Future precipitation is generated recursively in latent space from the terminal historical state. After the context $(\mathbf{H}_t,\mathbf{C}_t)$ has been formed, let $\mathbf{U}_{t+k}$ denote the recurrent input during future prediction:

\begin{equation}
\mathbf{U}_{t+k}
=
\left\{
\renewcommand{\arraystretch}{0.7}
\begin{array}{ll}
\mathbf{0}, & k=1,\\
\mathbf{H}_{t+k-1}, & k=2,\ldots,K.
\end{array}
\right.
\label{eq:future_input}
\end{equation}

where $\mathbf{0}$ denotes a zero tensor with the same spatial and channel dimensions as the latent recurrent input. The future recurrent states are then generated using the same ConvLSTM transition as in the historical context:

\begin{equation}
(\mathbf{H}_{t+k},\mathbf{C}_{t+k})
=
F_{\mathrm{CLSTM}}
\left(
\mathbf{U}_{t+k},
\mathbf{H}_{t+k-1},
\mathbf{C}_{t+k-1}
\right),
\qquad k=1,\ldots,K.
\label{eq:forecast_recurrence}
\end{equation}

Thus, the first forecast step uses a zero tensor as the recurrent input, while subsequent steps use the preceding hidden state. The final observed latent field $\mathbf{Z}_t$ is not reintroduced after the historical context has been constructed.

Each future hidden state is decoded to the transformed precipitation field,
\begin{equation}
\widehat{Y}_{t+k}^{*}=f_{\mathrm{reg}}\left(\mathbf{H}_{t+k};\theta_{\mathrm{reg}}\right).
\label{eq:regression_decoder}
\end{equation}
where $\theta_{\mathrm{reg}}$ denotes the trainable parameters of the precipitation regression decoder. The decoder applies a $3\times3$ convolution, bilinear upsampling to the original grid, a second $3\times3$ convolution, and a final $1\times1$ convolution. Predictions in physical units are recovered by
\begin{equation}
\widehat{Y}_{t+k}(\mathbf{s})=
\max\left\{
\exp\left[\sigma_Y\widehat{Y}_{t+k}^{*}(\mathbf{s})+\mu_Y\right]-1,
0
\right\},
\label{eq:inverse_prediction}
\end{equation}
where $\mu_Y$ and $\sigma_Y$ are the training set mean and standard deviation of the transformed precipitation target. All forecast horizons share the same transition and decoder parameters.

\subsection{Occurrence-intensity-aware Learning Objective}
\label{sec:occurrence_intensity}

Hourly precipitation is dominated by dry and weak rainfall samples, while moderate and heavy events are much less frequent. To reduce the tendency of an unweighted regression loss to favor conservative predictions, the baseline objective combines robust intensity regression, threshold event detection, and rainfall occurrence classification:
\begin{equation}
\mathcal{L}_{\mathrm{base}}
=\mathcal{L}_{\mathrm{int}}
+\lambda_{\mathrm{CSI}}\mathcal{L}_{\mathrm{CSI}}
+\lambda_{\mathrm{occ}}\mathcal{L}_{\mathrm{occ}}.
\label{eq:base_loss}
\end{equation}

For observed precipitation $Y$ in $\mathrm{mm\,h^{-1}}$, the intensity weight is
\begin{equation}
w(Y)=
\begin{cases}
1.0, & Y<0.1,\\
1.2, & 0.1\leq Y<1,\\
2.0, & 1\leq Y<4,\\
4.0, & 4\leq Y<8,\\
6.0, & Y\geq8.
\end{cases}
\label{eq:intensity_weights}
\end{equation}
The weighted regression loss is
\begin{equation}
\mathcal{L}_{\mathrm{int}}
=\frac{
\sum_{i,k}M_{i,k}w(Y_{i,k})
\rho_{\delta}(\widehat{Y}^{*}_{i,k}-Y^{*}_{i,k})
}{
\sum_{i,k}M_{i,k}w(Y_{i,k})+\epsilon
},
\label{eq:intensity_loss}
\end{equation}
where $M_{i,k}$ is the validity mask and $\rho_{\delta}$ denotes the Huber loss with $\delta=1.0$, applied after logarithmic transformation and normalization of precipitation. The intensity weights are normalized over valid observations during training, and $\epsilon$ is a small constant used to avoid division by zero.

To directly encourage threshold event detection, the hard event indicator
$I(Y\geq\tau)$ is paired with a smooth predicted event probability,

\begin{equation}
q_{\tau}(\widehat{Y})
=
\sigma\left[a(\widehat{Y}-\tau)\right].
\label{eq:soft_event_probability}
\end{equation}

Here, $a=3.0$ (mm h$^{-1}$)$^{-1}$ controls the sharpness of the smooth
transition around threshold $\tau$.

The differentiable CSI at horizon $k$ and threshold $\tau$ is written compactly as
\begin{equation}
\mathrm{CSI}^{\mathrm{soft}}_{k,\tau}
=
\frac{
\sum_i M_{i,k}q_{\tau}(\widehat{Y}_{i,k})I(Y_{i,k}\geq\tau)
}{
\sum_i M_{i,k}\left[q_{\tau}(\widehat{Y}_{i,k})+I(Y_{i,k}\geq\tau)-q_{\tau}(\widehat{Y}_{i,k})I(Y_{i,k}\geq\tau)\right]+\epsilon
}.
\label{eq:soft_csi}
\end{equation}
The corresponding event loss averages $1-\mathrm{CSI}^{\mathrm{soft}}_{k,\tau}$ over forecast horizons and the thresholds $\tau=1,4,8\ \mathrm{mm\,h^{-1}}$, with threshold weights $0.2$, $0.5$, and $0.3$, respectively. Horizon Weights are determined based on different experimental settings. Four-hour experiment uses horizon weights $(0.8,0.9,1.1,1.2)$, and to ensure the stability of the eight hour experiment, horizon weights are applied only to the final four time steps.

A parallel event decoder additionally predicts the occurrence of rainfall in $\tau_r=0.1\ \mathrm{mm\,h^{-1}}$,
\begin{equation}
R_{t+k}(\mathbf{s})=I\left[Y_{t+k}(\mathbf{s})\geq\tau_r\right],
\qquad
\widehat{p}_{t+k}(\mathbf{s})
=P\left[R_{t+k}(\mathbf{s})=1\mid\mathcal{H}_t,\mathbf{g}(\mathbf{s})\right],
\label{eq:event_probability}
\end{equation}
and $\mathcal{L}_{\mathrm{occ}}$ is the corresponding binary cross-entropy loss. The regression and event heads share the same latent states but use separate output parameters. We use this occurrence-intensity formulation as an auxiliary multi task objective rather than a strict hurdle model. In the baseline configuration, $\lambda_{\mathrm{CSI}}=\lambda_{\mathrm{occ}}=0.05$.

\subsection{Physics-guided fine-tuning through water vapor-budget constraints}
\label{sec:physics_guided}

The data-driven objective in Section~\ref{sec:occurrence_intensity} improves learning of precipitation intensity and occurrence but does not explicitly constrain the forecasts using information about atmospheric moisture availability. We therefore apply a second stage of physics-guided fine-tuning. The baseline architecture and $\mathcal{L}_{\mathrm{base}}$ remain unchanged, while moisture budget information is introduced only as a weak regularizer during training and is not used as an additional predictor during inference.

For an hourly interval, a precipitation-support diagnostic is derived from the vertically integrated water vapor budget,
\begin{equation}
P_{\mathrm{b}}=E-\Delta W-D_q,
\qquad
\Delta W=W_t-W_{t-1},
\label{eq:water_budget}
\end{equation}
where $E$ is upward positive evaporation, $W$ is total column water vapor, and $D_q$ is vertically integrated moisture-flux divergence. Only the positive part is used as precipitation evidence,
\begin{equation}
\widetilde{P}_{\mathrm{b}}
=\alpha\max(P_{\mathrm{b}},0),
\qquad
\alpha=
\frac{\overline{Y}_{\mathrm{train}}}
{\overline{\max(P_{\mathrm{b}},0)}_{\mathrm{train}}}.
\label{eq:calibrated_budget}
\end{equation}
The diagnostic is therefore treated as a soft physical indicator rather than an exact pixelwise precipitation target. A training set lower offset $q_{\mathrm{low}}$ defines one-sided regional lower bounds, so only physically unsupported underprediction is penalized; no physical upper bound or deterministic dry region penalty is imposed.

\begin{figure}[!htbp]
    \centering
    \includegraphics[width=01\textwidth]{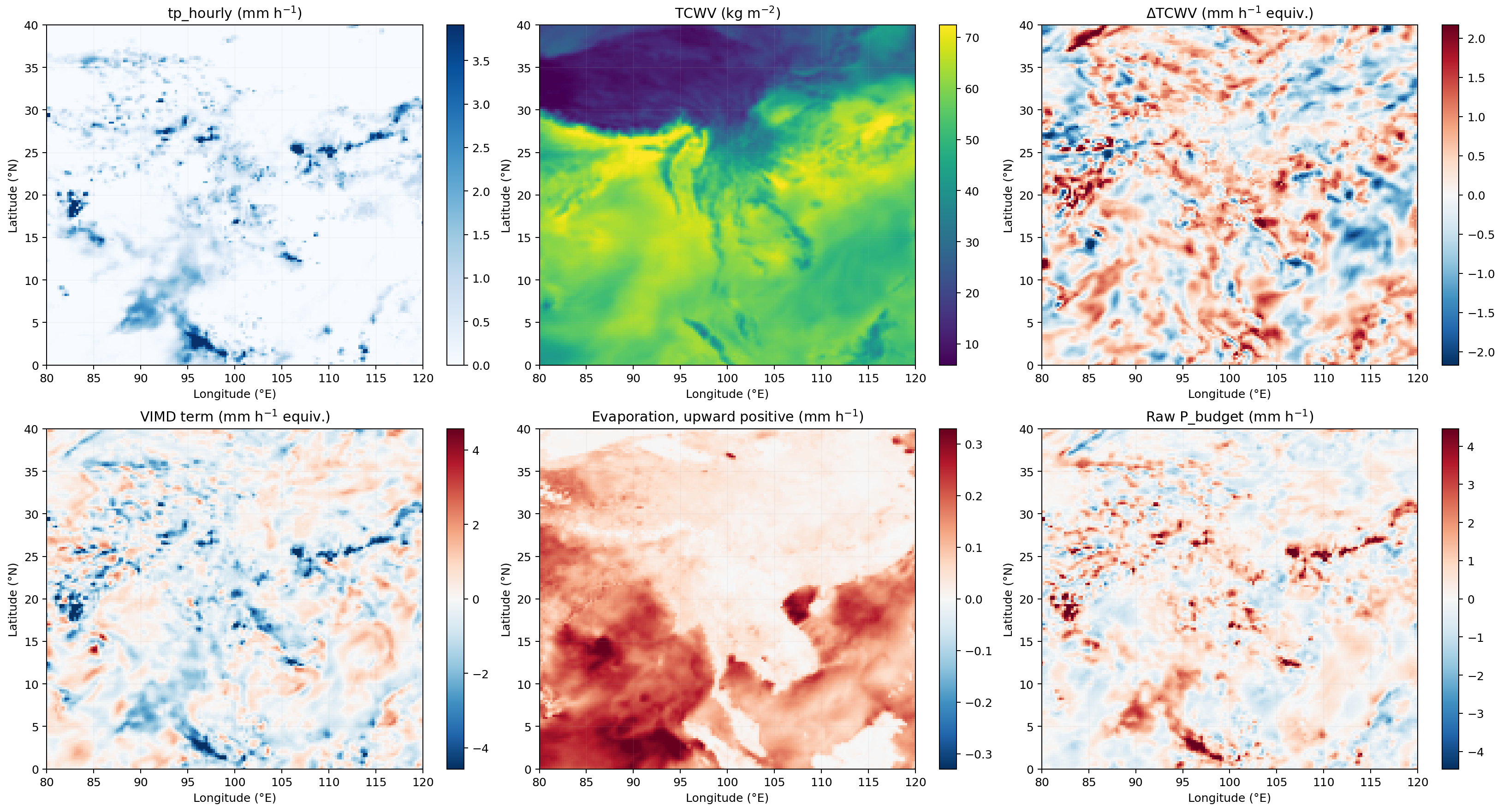}
    \caption{Spatial patterns of variables in the moisture budget equation at a specific time.}
    \label{fig:Raw budget}
\end{figure}

\begin{figure}[!htbp]
    \centering
    \includegraphics[width=0.75\textwidth]{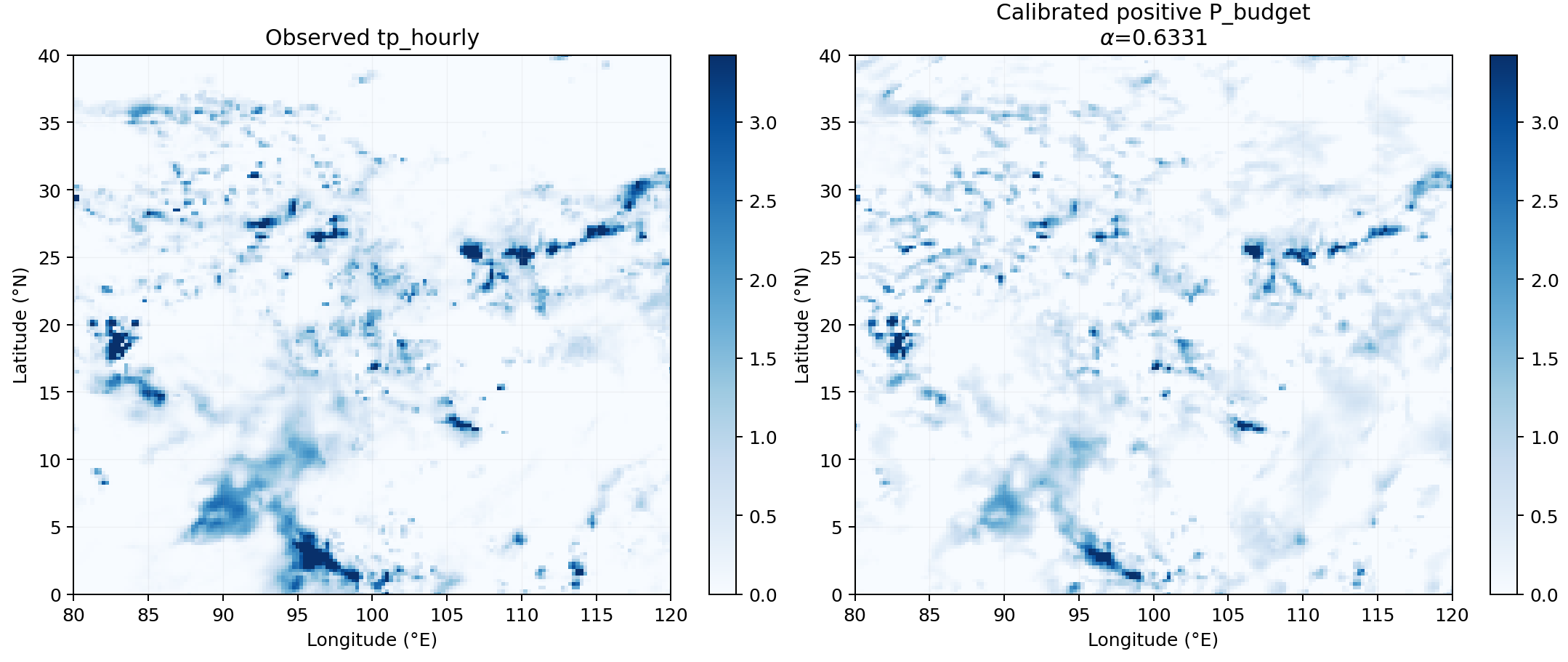}
    \caption{Spatial characteristics of the computed moisture budget physical constraints based on the formula.}
    \label{fig:calibrated budget}
\end{figure}

Figures~\ref{fig:Raw budget} and ~\ref{fig:calibrated budget} show the spatial distributions of the variables in Equations~\eqref{eq:water_budget} and ~\eqref{eq:calibrated_budget}, from which we can broadly estimate the spatial structures and magnitudes through calculation and calibration.

Three complementary constraints are constructed to provide physical guidance from different perspectives: local precipitation event occurrence, regional precipitation magnitude, and cumulative precipitation consistency. Let $\mathcal{K}_{\mathrm{phys}}\subseteq\{1,\ldots,K\}$ denote the set of forecast lead times at which physical guidance is activated. The choice of $\mathcal{K}_{\mathrm{phys}}$ is treated as part of the experimental configuration and can be adapted to the prediction task and data regime, rather than being fixed in the general formulation.

First, a neighborhood-based event constraint is introduced to account for spatial displacement between the water budget diagnostic and the predicted precipitation field. For each grid location $\mathbf{s}$, the maximum predicted precipitation within a local neighborhood is defined as

\begin{equation}
\widehat{Y}^{\mathrm{nbr}}_{t+k}(\mathbf{s})
=
\max_{\mathbf{u}\in\mathcal{N}_{3}(\mathbf{s})}
\widehat{Y}_{t+k}(\mathbf{u}),
\label{eq:neighborhood_prediction}
\end{equation}
where $\mathcal{N}_{3}(\mathbf{s})$ denotes the $3\times3$ neighborhood centered at $\mathbf{s}$. For precipitation thresholds $\tau\in\{4,8\}$ mm h$^{-1}$, the physical event indicator and the corresponding soft event probability are defined as

\begin{equation}
I^{\mathrm{phys}}_{k,\tau}(\mathbf{s})
=
I
\left[
\widetilde{P}_{b,t+k}(\mathbf{s})
\geq\tau
\right],
\end{equation}

\begin{equation}
p^{\mathrm{nbr}}_{k,\tau}(\mathbf{s})
=
\sigma
\left(
\frac{
\widehat{Y}^{\mathrm{nbr}}_{t+k}(\mathbf{s})-\tau
}{b}
\right),
\end{equation}
where $\sigma(\cdot)$ denotes the sigmoid function and $b=0.75$ mm h$^{-1}$ is a softness parameter controlling the transition around the precipitation threshold. The neighborhood event loss is calculated using physically supported event locations over the selected physical-guidance lead times:

\begin{equation}
\mathcal{L}_{\mathrm{nbr},\tau}
=
-
\frac{
\displaystyle
\sum_{k\in\mathcal{K}_{\mathrm{phys}}}
\sum_{\mathbf{s}}
I^{\mathrm{phys}}_{k,\tau}(\mathbf{s})
\log
p^{\mathrm{nbr}}_{k,\tau}(\mathbf{s})
}
{
\displaystyle
\sum_{k\in\mathcal{K}_{\mathrm{phys}}}
\sum_{\mathbf{s}}
I^{\mathrm{phys}}_{k,\tau}(\mathbf{s})
+\epsilon
}.
\end{equation}

The moderate- and heavy-precipitation constraints are then combined as

\begin{equation}
\mathcal{L}_{\mathrm{nbr}}
=
0.6\mathcal{L}_{\mathrm{nbr},4}
+
0.4\mathcal{L}_{\mathrm{nbr},8}.
\end{equation}

Second, a multiscale regional lower-bound constraint is introduced to avoid excessive precipitation attenuation at larger spatial scales. For a spatial scale $m\in\{4,8\}$, the precipitation fields are aggregated into corresponding spatial patches:

\begin{equation}
\overline{\widehat{Y}}^{(m)}_{t+k,r}
=
\mathrm{mean}
\left(
\widehat{Y}_{t+k}
\right),
\qquad
\overline{\widetilde{P}}^{(m)}_{b,t+k,r}
=
\mathrm{mean}
\left(
\widetilde{P}_{b,t+k}
\right),
\end{equation}
where $r$ denotes the patch index. The physically supported lower bound is defined as

\begin{equation}
B^{(m)}_{t+k,r}
=
\max
\left(
\overline{\widetilde{P}}^{(m)}_{b,t+k,r}
+
q_{\mathrm{low}},
0
\right),
\end{equation}
where $q_{\mathrm{low}}$ is a training-derived calibration offset estimated from the discrepancy between observed precipitation and the water budget diagnostic. Only underestimation below this physical lower bound is penalized. For each valid patch, the one-sided violation is

\begin{equation}
\ell^{(m)}_{k,r}
=
\left[
\log
\left(1+B^{(m)}_{t+k,r}\right)
-
\log
\left(1+\overline{\widehat{Y}}^{(m)}_{t+k,r}\right)
\right]_{+}.
\end{equation}

The regional loss on the spatial scale $m$ is obtained by averaging these violations over the selected lead times and valid spatial patches.

\begin{equation}
\mathcal{L}_{m}
=
\frac{
\displaystyle
\sum_{k\in\mathcal{K}_{\mathrm{phys}}}
\sum_{r\in\mathcal{R}_{m}(k)}
\ell^{(m)}_{k,r}
}
{
\displaystyle
\sum_{k\in\mathcal{K}_{\mathrm{phys}}}
|\mathcal{R}_{m}(k)|
+\epsilon
},
\end{equation}
where $\mathcal{R}_{m}(k)$ denotes the set of valid patches at the lead time $k$. The multiscale regional constraint is then formulated as

\begin{equation}
\mathcal{L}_{\mathrm{ms}}
=
0.6\mathcal{L}_{4}
+
0.4\mathcal{L}_{8}.
\end{equation}

Here, $\mathcal{L}_{4}$ and $\mathcal{L}_{8}$ represent regional constraints computed from $4\times4$ and $8\times8$ spatial patches, respectively.

Third, a cumulative consistency constraint can be introduced to reduce excessive attenuation across a selected group of forecast lead times. Let $\mathcal{K}_{\mathrm{cum}}\subseteq\mathcal{K}_{\mathrm{phys}}$ denote the lead times included in the cumulative constraint. The corresponding accumulated precipitation fields are defined as

\begin{equation}
\widehat{Y}_{\mathrm{cum}}
=
\sum_{k\in\mathcal{K}_{\mathrm{cum}}}
\widehat{Y}_{t+k},
\qquad
P_{\mathrm{cum}}
=
\sum_{k\in\mathcal{K}_{\mathrm{cum}}}
\widetilde{P}_{b,t+k}.
\end{equation}

The same multiscale one-sided lower-bound formulation is then applied to the accumulated precipitation fields, yielding the cumulative constraint $\mathcal{L}_{\mathrm{cum}}$. This formulation allows precipitation to be redistributed among the selected forecast hours while discouraging excessive attenuation of their accumulated amount.

Finally, the complete physical regularization term is expressed as

\begin{equation}
\mathcal{L}_{\mathrm{phys}}
=
0.25\mathcal{L}_{\mathrm{nbr}}
+
0.35\mathcal{L}_{\mathrm{ms}}
+
0.40\mathcal{L}_{\mathrm{cum}},
\label{eq:physical_loss}
\end{equation}

and the fine-tuning objective is

\begin{equation}
\mathcal{L}_{FT}
=
\mathcal{L}_{base}
+
\lambda_{\mathrm{phys}}
R(e)
\mathcal{L}_{\mathrm{phys}},
\label{eq:fine_tuning_loss}
\end{equation}

where $R(e)$ controls the gradual introduction of physical regularization during fine-tuning. Therefore, the physical information acts as a weak asymmetric constraint that discourages physically unsupported precipitation attenuation while preserving the flexibility of the data-driven forecast.

\section{Experimental Design}
\label{sec:experimental-design}

The experimental design is constructed to separate the effects of model architecture,  event-aware learning, forecast horizon, and physics-guided fine-tuning. Three complementary settings are considered. The regional domain experiment provides an initial comparison between the reconstructed GENESIS$_{\mathrm{FL}}$ benchmark and several variants of the proposed geographically explicit ConvLSTM framework. The enlarged domain experiment then evaluates the proposed approach over a substantially broader spatial and temporal sample, first under a common 4-h forecasting task and subsequently under an extended 8-h forecasting horizon. Finally, spatial subsets of the enlarged domain are used to investigate whether the contribution of the moisture budget constraint changes when the available spatial information is more restricted.

\subsection{Data and preprocessing}
\label{sec:data-preprocessing}

All experiments use hourly ERA5 atmospheric fields~\citep{hersbach2020era5}. The regional experiment focuses on East Asia ($22^\circ$--$32^\circ$N and $110^\circ$--$120^\circ$E) with a $0.1^\circ$ resolution, spanning July 2020--2025. Six dynamic predictors are used in this setting: total precipitation (\texttt{tp}), 10-m zonal wind (\texttt{u10}), 10-m meridional wind (\texttt{v10}), 2-m dew-point temperature (\texttt{d2m}), 2-m temperature (\texttt{t2m}), and surface pressure (\texttt{sp}). Data from 2020-2023 are used for training, 2024 for validation, and 2025 for testing.

The enlarged domain experiment covers $0^{\circ}$--$40^{\circ}$N and $80^{\circ}$--$120^{\circ}$E  (South, Southeast, and East Asia) during April--September of 2020-2023 at a $0.25^\circ$ resolution. To obtain a more parsimonious predictor configuration over the substantially larger domain, we retain four dynamic variables: \texttt{tp}, \texttt{u10}, \texttt{v10}, and \texttt{d2m}. We use 2020--2021 for training, 2022 for validation, and 2023 for testing. The spatial subset experiments use the same temporal period, predictor set, and train--validation--test partition as the enlarged domain experiment, but restrict the forecasting domain to approximately one-sixteenth of the enlarged spatial region. For the detailed subset experiment, the central portion of the enlarged domain is selected by reducing both the latitudinal and longitudinal extents to one quarter of their original ranges, yielding an area of approximately $1/16$; for the robustness analysis, the full enlarged domain is partitioned into a $4\times4$ grid of 16 non-overlapping spatial subsets of the same size.

As described in Section~\ref{sec:problem_formulation}, precipitation is transformed using $\log(1+Y)$ prior to model fitting. The transformed precipitation field and the remaining dynamic predictors are standardized using statistics computed from the training set only; the same transformations are subsequently applied to the validation and test sets. Static geographic features are constructed directly from latitude and longitude according to the representation introduced in Section~3.2.

\subsection{Experimental configurations}
\label{sec:experimental-configurations}

Table~\ref{tab:experimental-settings} summarizes the three experimental settings. The regional experiment uses a six-hour historical input sequence to predict the next four hourly precipitation fields. This setting is used to compare the reconstructed GENESIS$_{\mathrm{FL}}$ model with the baseline and enhanced variants of the proposed framework.

The enlarged domain experiments serve two purposes. First, a common 4-h forecasting task is used to compare GENESIS$_{\mathrm{FL}}$ and the proposed event-aware model under the same forecast horizon. Second, the forecasting horizon is extended to eight hours to evaluate degradation in predictive skill and to examine the contribution of physics-guided fine-tuning at longer lead times. In this main experiment, eight historical hourly fields are used to predict the subsequent eight precipitation fields.

Finally, the spatial subset experiment retains the 8-h input--output configuration while restricting the spatial domain. This experiment is designed to examine whether physical regularization becomes more influential when the data-driven model has access to less spatial context.

\begin{table}[htbp]
\centering
\caption{Summary of the experimental settings.}
\label{tab:experimental-settings}
\begin{tabular}{
  >{\raggedright\arraybackslash}p{2.0cm}
  >{\raggedright\arraybackslash}p{3.6cm}
  >{\raggedright\arraybackslash}p{2.3cm}
  >{\raggedright\arraybackslash}p{2.8cm}
  >{\raggedright\arraybackslash}p{3.2cm}
}
\hline
Experiment &
Domain and period &
Predictors &
Train / validation / test &
Forecast task and comparison \\
\hline

Regional &
$22^{\circ}$--$32^{\circ}$N,
$110^{\circ}$--$120^{\circ}$E;
each July 2020--2025 &
\texttt{tp}, \texttt{u10}, \texttt{v10},
\texttt{d2m}, \texttt{t2m}, \texttt{sp} &
2020--2023 / 2024 / 2025 &
6-h $\rightarrow$ 4-h;
GENESIS$_{\mathrm{FL}}$, ConvLSTM,
GCL, GCL$_E$, GCL$_P$ \\

Enlarged &
$0^{\circ}$--$40^{\circ}$N,
$80^{\circ}$--$120^{\circ}$E;
April--September 2020--2023 &
\texttt{tp}, \texttt{u10}, \texttt{v10}, \texttt{d2m} &
2020--2021 / 2022 / 2023 &
6-h $\rightarrow$ 4-h benchmark;
8-h $\rightarrow$ 8-h main task;
GENESIS$_{\mathrm{FL}}$ vs.\ GCL$_E$,
and GCL$_E$ vs.\ GCL$_P$ \\

Spatial subset &
Approximately $1/16$ of the enlarged domain;
same temporal period &
\texttt{tp}, \texttt{u10}, \texttt{v10}, \texttt{d2m} &
Same as enlarged &
8-h $\rightarrow$ 8-h;
GCL$_E$ vs.\ GCL$_P$ \\

\hline
\end{tabular}
\end{table}

\subsection{Comparison models and ablation strategy}
\label{sec:model-comparisons}

The proposed framework is evaluated through a sequence of increasingly structured model variants designed to isolate the contribution of its main components. ConvLSTM denotes the conventional convolutional recurrent baseline without the proposed geographic representation. GCL augments the convolutional recurrent architecture with the explicit geographic embedding described in Section~3.2 while retaining a standard data-driven training objective. GCL$_E$ further incorporates the occurrence-intensity-aware objective of Section~3.5, which combines weighted robust regression, differentiable event-oriented learning, and auxiliary precipitation-occurrence prediction. Finally, GCL$_P$ denotes the physics-guided model obtained by fine-tuning GCL$_E$ using the atmospheric water budget regularization introduced in Section~3.6.

This sequence provides a natural component-wise comparison: ConvLSTM $\rightarrow$ GCL evaluates the contribution of explicit geographic information, GCL $\rightarrow$ GCL$_E$ evaluates the effect of occurrence-intensity-aware learning, and GCL$_E \rightarrow$ GCL$_P$ evaluates the incremental effect of the physics-guided fine-tuning stage. The regional experiment includes the full sequence of models, whereas the larger experiments focus on the comparisons most relevant to robustness across spatial settings and longer lead forecasting.

GENESIS$_{\mathrm{FL}}$ is included as an event-oriented ConvLSTM benchmark. Because the original training dataset and several implementation details of the published system are unavailable, GENESIS$_{\mathrm{FL}}$ is a reconstructed implementation rather than an exact reproduction of the originally reported model. It is therefore evaluated under the same ERA5-based data partitions and forecasting protocol as the proposed models, and the comparison should be interpreted as an architectural benchmark rather than as an attempt to reproduce the numerical results reported in the original study.

\subsection{Training and implementation}
\label{sec:training-details}

All model selection decisions are based exclusively on the training and validation sets, while the test sets are reserved for final evaluation. The same preprocessing statistics are used across training, validation, and testing within each experimental setting. For matched comparisons, competing model variants use the same input sequences, forecast horizons, and train--validation--test partitions.

GCL$_E$ is first optimized using the occurrence-intensity-aware objective $\mathcal{L}_{\mathrm{base}}$ defined in Section~3.5. The physics-guided model GCL$_P$ is initialized from the corresponding trained GCL$_E$ model and subsequently fine-tuned using
\[
\mathcal{L}_{\mathrm{FT}} = \mathcal{L}_{\mathrm{base}}
+
\lambda_{\mathrm{phys}} R(e)\mathcal{L}_{\mathrm{phys}}, \] as defined in Section~3.6. Thus, the water budget information modifies the model only during the second-stage fine-tuning procedure and is not provided as an additional predictor during inference.

Models are optimized using the Adam optimizer with an initial learning rate of \(5\times10^{-4}\). We set the batch size to 16, and each baseline model is trained for up to 100 epochs. Early stopping is applied with a patience of 8 epochs. The learning rate is adjusted using ReduceLROnPlateau, with a reduction factor of 0.5 and a patience of 3 epochs. A weight decay of 0 is used. All experiments use random seed 42.

For GCL$_P$, physics-guided fine-tuning is initialized from the best GCL$_E$ checkpoint and uses a reduced learning rate of \(5\times10^{-5}\). The physics-guided fine-tuning coefficient is set to \(\lambda_{\mathrm{phys}}=0.02\), and fine-tuning is performed for up to 20 epochs, with an early-stopping patience of 8 epochs. During fine-tuning, ReduceLROnPlateau is used with a reduction factor of 0.5 and a patience of 2 epochs. The ramp function \(R(e)\) gradually introduces the physical regularization over the first 5 fine-tuning epochs according to

$$
R(e)=\min\left(\frac{e}{5},1\right),
$$

where \(e\) denotes the fine-tuning epoch starting from 1. The calibration offset \(q_{\mathrm{low}}\) is estimated independently for each experimental spatial domain from all valid $4\times4$ patch residuals in the training years, using the 0.10 quantile of the discrepancy between observed precipitation and the calibrated water budget diagnostic. The same offset is used across forecast horizons.

Training is performed using Quadro GV100 GPUs, each with 32 GB of memory. The main GCL$_E$ model requires approximately 8 hours for training, while the GCL$_P$ fine-tuning stage requires approximately 6 hours.

\subsection{Evaluation metrics}
\label{sec:evaluation-metrics}

Forecast performance is evaluated using both continuous and categorical metrics because precipitation prediction requires accurate estimation of rainfall magnitude as well as successful detection of comparatively rare moderate and heavy precipitation events.

Continuous predictive accuracy is assessed using mean absolute error (MAE), root mean squared error (RMSE), mean bias, and Pearson correlation. MAE and RMSE quantify the overall magnitude of forecast errors, bias measures systematic over- or underprediction, and correlation evaluates the correspondence between predicted and observed spatial variability.

Event-based performance is evaluated at precipitation thresholds of $1$, $4$, and $8$ mm h$^{-1}$. The critical success index (CSI) measures the fraction of observed or predicted events that are correctly detected, while the Heidke skill score (HSS) assesses event prediction relative to chance. The false alarm ratio (FAR) quantifies the fraction of predicted events that are not observed. Probability of detection (POD) is additionally reported in experiments designed to examine the long-lead contribution of physics-guided fine-tuning. Among these metrics, CSI and HSS provide relatively comprehensive assessments of event-based forecast performance by accounting for both successful detections and forecast errors.

The $4$ and $8$ mm h$^{-1}$ thresholds receive particular attention because they represent the moderate- and heavy-precipitation regimes explicitly emphasized by the event-aware and physics-guided components of the proposed framework. Metrics are reported both in aggregate and, where relevant, separately by forecast lead time to determine how predictive skill and the contribution of physical guidance evolve as the forecast horizon increases.

\section{Results}
\label{sec:results}

The experimental results are presented in three stages. We first examine the regional experiment, where the sequence ConvLSTM $\rightarrow$ GCL $\rightarrow$ GCL$_E$ $\rightarrow$ GCL$_P$ provides a component-wise assessment of geographic representation, event-aware learning, and physics-guided fine-tuning. We then evaluate the proposed framework over the enlarged domain under both the common 4-h forecasting task and the extended 8-h setting. Finally, the spatial subset experiments are used to examine the behavior of the physics-guided constraint across more spatially restricted forecasting regimes.

\subsection{Regional comparison}
\label{sec:regional-results}

Table~\ref{tab:regional-results} summarizes the overall forecasting performance on the regional ERA5 dataset. The results reveal a clear distinction between improvements in continuous pointwise accuracy and improvements in precipitation event detection.

\begin{table}[!htbp]
\centering
\caption{Overall forecasting performance on the regional ERA5 dataset (best value in {\bf bold}).}
\label{tab:regional-results}
\resizebox{\textwidth}{!}{
\begin{tabular}{lrrrrrrrrrrrrr}
\hline
Model &
MAE & RMSE & Bias & Corr &
CSI@1 & CSI@4 & CSI@8 &
HSS@1 & HSS@4 & HSS@8 &
FAR@1 & FAR@4 & FAR@8 \\
\hline
GENESIS$_{\mathrm{FL}}$ &
\textbf{0.2172} & 0.6583 & -0.0771 & 0.6043 &
0.2749 & 0.0888 & 0.0275 &
0.4016 & 0.1611 & 0.0534 &
0.3603 & \textbf{0.2976} & \textbf{0.2505} \\

ConvLSTM &
0.2302 & 0.6633 & -0.0599 & 0.5960 &
0.2700 & 0.0752 & 0.0324 &
0.3968 & 0.1380 & 0.0626 &
\textbf{0.3360} & 0.3379 & 0.4675 \\

GCL &
0.2289 & \textbf{0.6407} & -0.0297 & \textbf{0.6177} &
0.3234 & 0.1123 & 0.0462 &
0.4546 & 0.1991 & 0.0880 &
0.4281 & 0.4002 & 0.4237 \\

GCL$_E$ &
0.2556 & 0.6609 & 0.0314 & 0.6053 &
0.3385 & 0.1524 & 0.1005 &
0.4641 & 0.2600 & 0.1821 &
0.5227 & 0.5525 & 0.5401 \\

GCL$_P$ &
0.2466 & 0.6676 & \textbf{0.0172} & 0.6143 &
\textbf{0.3420} & \textbf{0.1865} & \textbf{0.1718} &
\textbf{0.4712} & \textbf{0.3090} & \textbf{0.2923} &
0.4859 & 0.5774 & 0.6174 \\
\hline
\end{tabular}
}
\end{table}

\subsubsection{Contribution of geographic and event-aware components}

The comparison between ConvLSTM and GCL isolates the effect of explicitly incorporating geographic information. Adding the geographic representation reduces RMSE from 0.6633 to 0.6407 and increases correlation from 0.5960 to 0.6177. Improvements are also observed for the event-based metrics: CSI increases from 0.2700, 0.0752, and 0.0324 for ConvLSTM to 0.3234, 0.1123, and 0.0462 for GCL at the 1, 4, and 8 mm h$^{-1}$ thresholds, respectively. These results indicate that supplementing convolutional spatiotemporal features with absolute geographic information improves both overall representation of the precipitation field and detection of precipitation events.

Introducing the occurrence-intensity-aware objective produces a different effect. Compared with GCL, GCL$_E$ substantially improves event-oriented performance, with CSI@4 increasing from 0.1123 to 0.1524 and CSI@8 from 0.0462 to 0.1005. Corresponding HSS values increase from 0.1991 to 0.2600 at 4 mm h$^{-1}$ and from 0.0880 to 0.1821 at 8 mm h$^{-1}$. These improvements are accompanied by higher MAE and RMSE and increased FAR. The event-aware objective therefore shifts the model away from conservative predictions dominated by weak precipitation and toward greater sensitivity to moderate and heavy precipitation events.

GCL$_P$ further strengthens this event-oriented behavior. It achieves the highest CSI and HSS at all three precipitation thresholds. The largest differences occur for heavy precipitation: CSI@8 increases from 0.1005 for GCL$_E$ to 0.1718 for GCL$_P$, while HSS@8 increases from 0.1821 to 0.2923. Relative to GENESIS$_{\mathrm{FL}}$, CSI@8 increases from 0.0275 to 0.1718 and HSS@8 from 0.0534 to 0.2923. At the same time, GCL$_P$ does not achieve the lowest pointwise error, showing that improved detection of rare high-intensity precipitation is not equivalent to minimizing average regression error.

\subsubsection{Lead-time dependence of event skill}

\begin{figure}[!htbp]
    \centering
    \includegraphics[width=\textwidth]{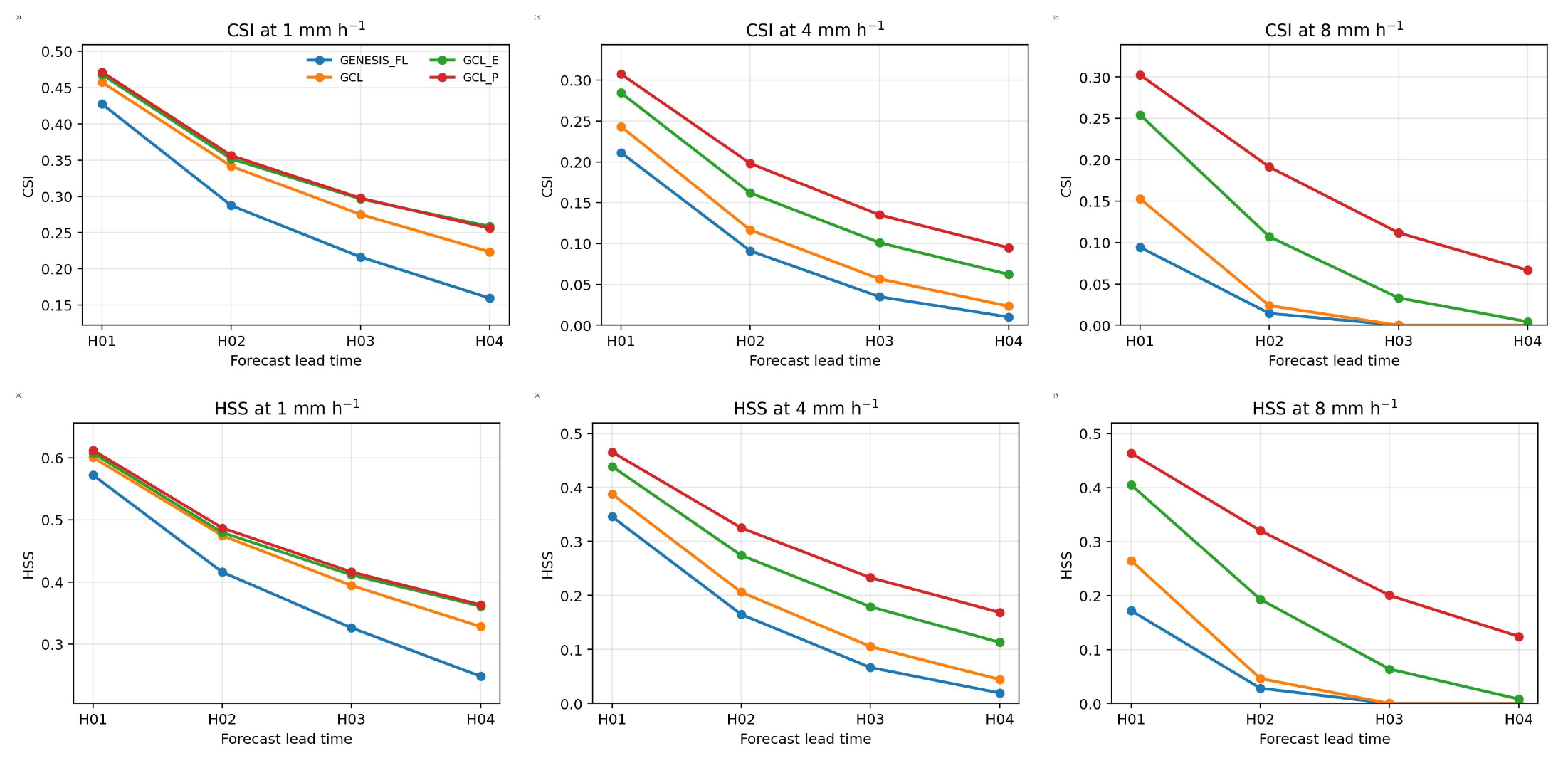}
    \caption{Lead-time-dependent CSI and HSS on the regional dataset at
    precipitation thresholds of 1, 4, and 8 mm h$^{-1}$.}
    \label{fig:regional-leadtime}
\end{figure}

The lead-time-dependent results further clarify these differences in Figure~\ref{fig:regional-leadtime}. Event skill decreases with increasing forecast horizon for all models, with the decline becoming substantially more pronounced as the precipitation threshold increases. Heavy precipitation at 8 mm h$^{-1}$ is therefore considerably more difficult to preserve at later lead times than weak or moderate precipitation.

The separation among the models is most apparent at the 4 and 8 mm h$^{-1}$ thresholds. At H03--H04, GCL$_P$ retains measurable heavy-precipitation skill after GENESIS$_{\mathrm{FL}}$ and GCL have nearly lost it. The behavior is consistent with the purpose of the event-aware and physics-guided components: rather than optimizing only average pointwise accuracy, these components reduce the rapid attenuation of signals associated with moderate and heavy precipitation during recursive forecasting.

\subsection{Enlarged-domain comparison}
\label{sec:enlarged-results}

The enlarged domain experiments evaluate whether the improvements observed in the regional setting remain evident over a substantially broader spatial and temporal sample. Two forecasting configurations are considered. The first uses a common 4-h forecasting task to provide a matched comparison with GENESIS$_{\mathrm{FL}}$. The second extends the forecasting horizon to 8 h and focuses on the incremental contribution of physics-guided fine-tuning.

\subsubsection{Common 4-h forecasting task}
\label{sec:enlarged-4h}

Table~\ref{tab:enlarged-4h} compares GENESIS$_{\mathrm{FL}}$ and GCL$_E$ under the
same 4-h forecasting task on the enlarged ERA5 dataset.

\begin{table}[!htbp]
\centering
\caption{Overall performance of GENESIS$_{\mathrm{FL}}$ and GCL$_E$ under the
common 4-h forecasting task on the enlarged ERA5 dataset.}
\label{tab:enlarged-4h}
\resizebox{\textwidth}{!}{
\begin{tabular}{lrrrrrrrrrrrrr}
\hline
Model &
MAE & RMSE & Bias & Corr &
CSI@1 & CSI@4 & CSI@8 &
HSS@1 & HSS@4 & HSS@8 &
FAR@1 & FAR@4 & FAR@8 \\
\hline
GENESIS$_{\mathrm{FL}}$ &
0.2220 & 0.6356 & -0.0434 & 0.6310 &
0.3208 & 0.1021 & 0.0675 &
0.4531 & 0.1830 & 0.1262 &
0.4041 & \textbf{0.3033} & \textbf{0.3020} \\

GCL$_E$-4h &
\textbf{0.1594} & \textbf{0.4616} & \textbf{-0.0027} & \textbf{0.7556} &
\textbf{0.4341} & \textbf{0.2909} & \textbf{0.2962} &
\textbf{0.5785} & \textbf{0.4481} & \textbf{0.4567} &
\textbf{0.3617} & 0.4114 & 0.4204 \\
\hline
\end{tabular}
}
\end{table}

Under this matched setting, GCL$_E$ substantially outperforms GENESIS$_{\mathrm{FL}}$ in both continuous and event-based metrics. MAE decreases from 0.2220 to 0.1594, corresponding to a reduction of approximately 28\%, while RMSE decreases from 0.6356 to 0.4616, a reduction of approximately 27\%. Correlation increases from 0.6310 to 0.7556, and the magnitude of the mean bias decreases from 0.0434 to 0.0027 mm h$^{-1}$.

The improvement is particularly pronounced for moderate and heavy precipitation. CSI@4 increases from 0.1021 to 0.2909, while CSI@8 increases from 0.0675 to 0.2962. Similar improvements are observed for HSS, which increases from 0.1830 to 0.4481 at 4 mm h$^{-1}$ and from 0.1262 to 0.4567 at 8 mm h$^{-1}$. These results show that the advantage of the geographically explicit and occurrence-intensity-aware model is not restricted to the initial regional experiment.

The FAR results nevertheless reveal the same sensitivity--false--alarm trade--off observed previously. GCL$_E$ reduces FAR at the 1 mm h$^{-1}$ threshold, but GENESIS$_{\mathrm{FL}}$ retains lower FAR at 4 and 8 mm h$^{-1}$. Thus, the large improvements in CSI and HSS for moderate and heavy precipitation are accompanied by a greater tendency to predict high intensity events that do not occur.

\subsubsection{Extended 8-h forecasting task}
\label{sec:enlarged-8h}

The extended 8-h experiment evaluates the incremental contribution of physics-guided fine-tuning after the data-driven GCL$_E$ model has already been trained. Table ~\ref{tab:enlarged-8h} reports the overall results.

\begin{table}[!htbp]
\centering
\caption{Overall performance of GCL$_E$ and GCL$_P$ under the extended 8-h
forecasting task on the enlarged ERA5 dataset.}
\label{tab:enlarged-8h}
\resizebox{\textwidth}{!}{
\begin{tabular}{lrrrrrrrrrrrrr}
\hline
Model &
MAE & RMSE & Bias & Corr &
CSI@1 & CSI@4 & CSI@8 &
HSS@1 & HSS@4 & HSS@8 &
FAR@1 & FAR@4 & FAR@8 \\
\hline
GCL$_E$-8h &
\textbf{0.1867} & \textbf{0.5315} & -0.0129 & \textbf{0.6586} &
0.3466 & 0.1984 & \textbf{0.1811} &
0.4838 & 0.3285 & \textbf{0.3063} &
\textbf{0.4231} & \textbf{0.4997} & \textbf{0.5360} \\

GCL$_P$-8h &
0.1909 & 0.5407 & \textbf{-0.0043} & 0.6509 &
\textbf{0.3479} & \textbf{0.1989} & 0.1795 &
\textbf{0.4845} & \textbf{0.3289} & 0.3039 &
0.4358 & 0.5392 & 0.5832 \\
\hline
\end{tabular}
}
\end{table}

In contrast to the common 4-h comparison, the incremental effect of physical regularization over the full enlarged domain is modest. GCL$_P$ reduces the magnitude of the negative bias from 0.0129 to 0.0043 mm h$^{-1}$ and produces small increases in CSI and HSS at the 1 and 4 mm h$^{-1}$ thresholds. However, GCL$_E$ retains slightly lower MAE and RMSE, higher correlation, and marginally higher CSI@8 and HSS@8. GCL$_P$ also produces larger FAR values at all three thresholds.

These results indicate that physics-guided fine-tuning does not provide a uniform improvement when applied over the full enlarged domain. Instead, its clearest aggregate effect in this setting is the reduction of systematic underprediction, while the remaining continuous and categorical metrics are nearly unchanged or slightly degraded.

\subsubsection{Spatial structure of the forecasts}

\begin{figure}[!htbp]
    \centering
    \includegraphics[width=\textwidth]{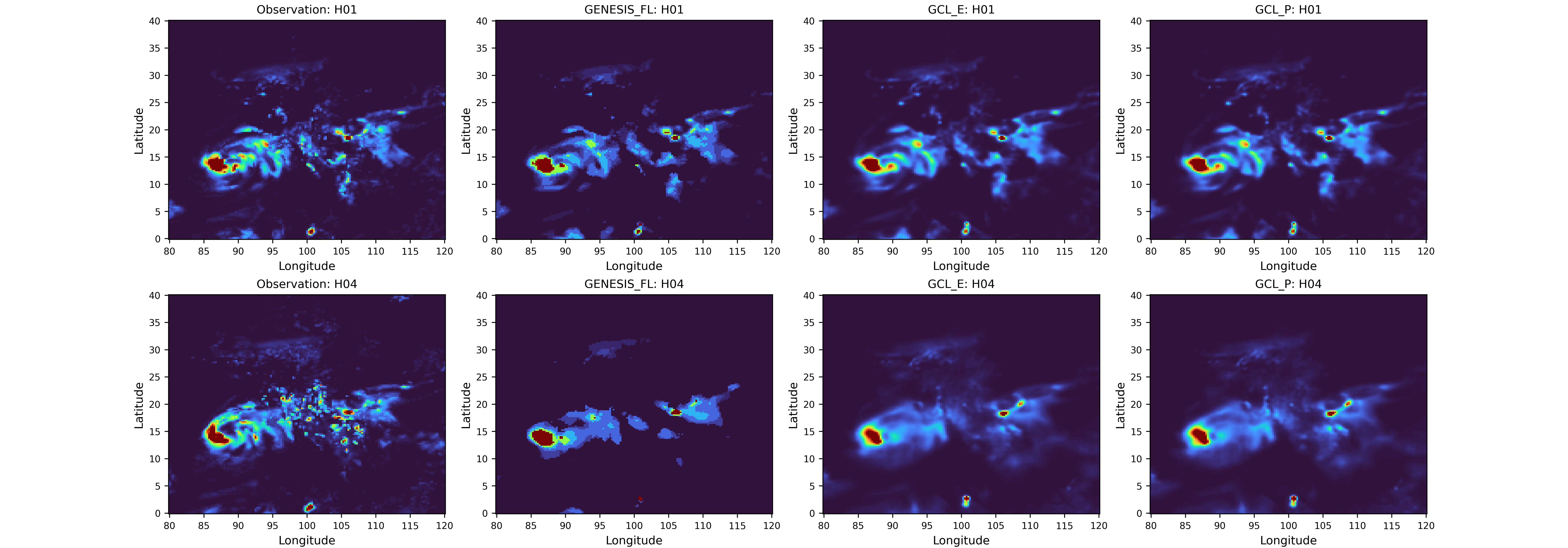}
    \hfill
    \includegraphics[width=\textwidth]{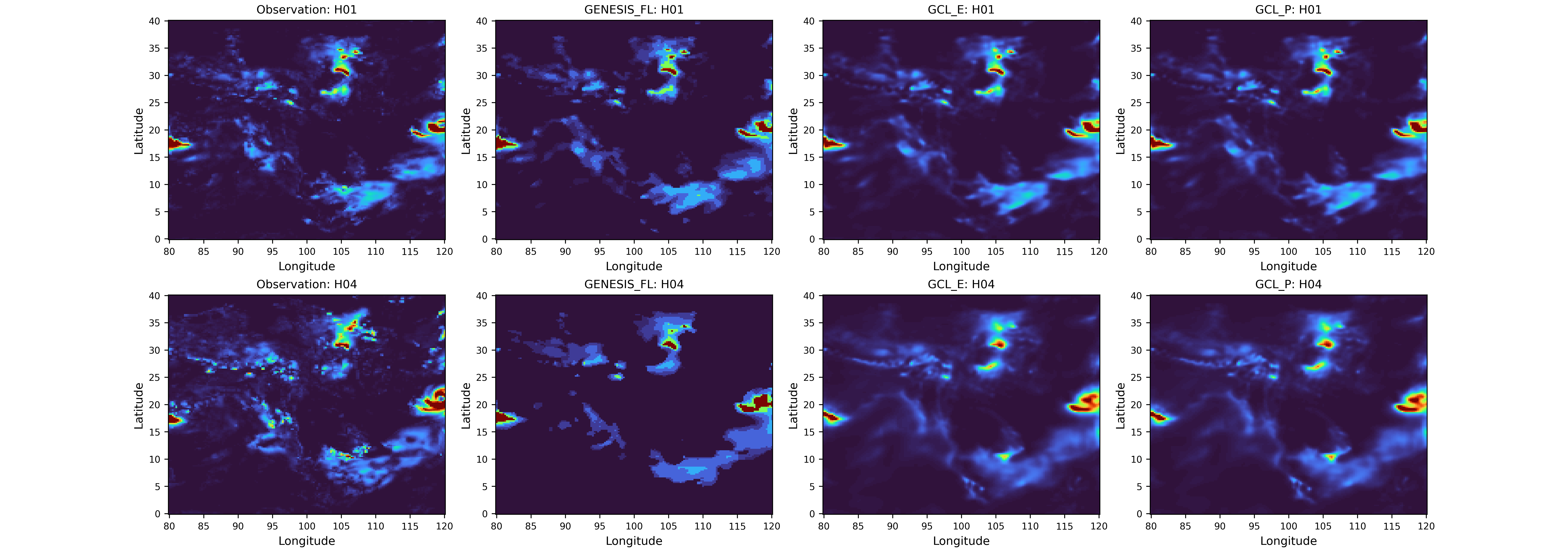}
    \caption{Representative precipitation forecasts for two groups on the enlarged ERA5 dataset at H01, H04.}
    \label{fig:enlarged-case}
\end{figure}

Figure~\ref{fig:enlarged-case} presents the predicted precipitation fields of different models at the same time step. At H01, all models capture the dominant precipitation systems, although the predicted fields are generally smoother than the observations. By H04, GCL$_E$ and GCL$_P$ preserve broader precipitation structures than GENESIS$_{\mathrm{FL}}$, consistent with their higher event-based skill at later forecast horizons.

\begin{figure}[!htbp]
    \centering
    \includegraphics[width=0.63\textwidth]{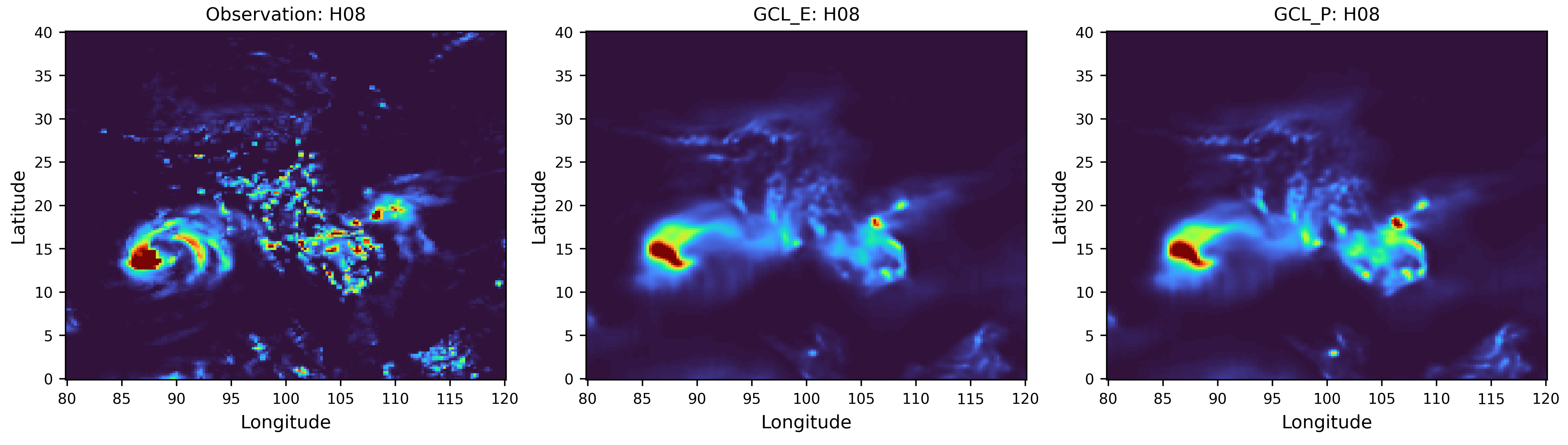}
    \hfill
    \includegraphics[width=0.63\textwidth]{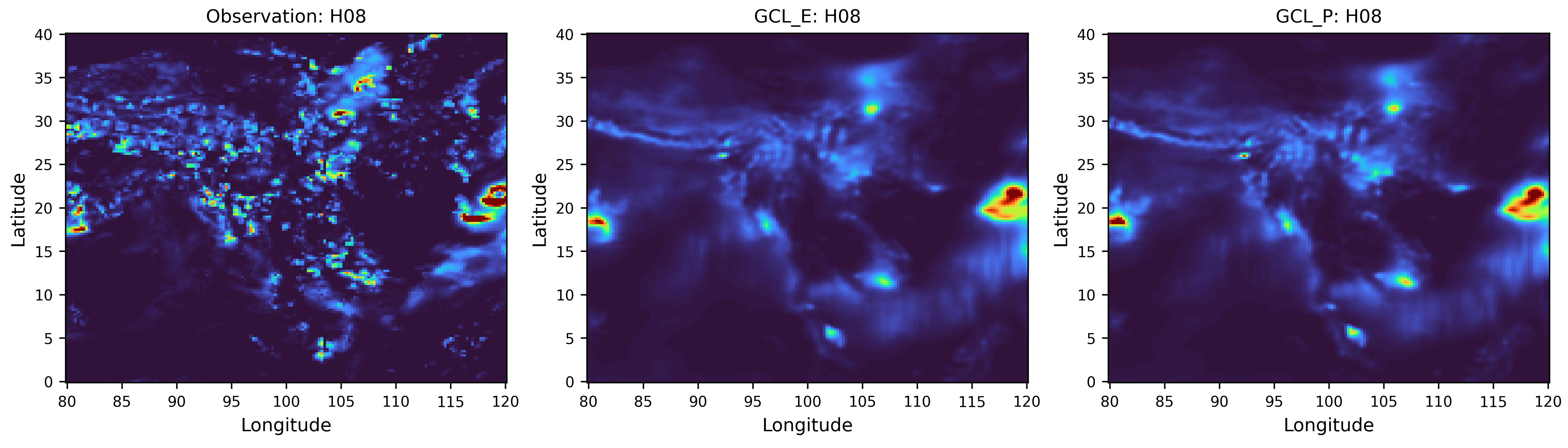}
    \caption{Representative precipitation forecasts for two cases at H08.}
    \label{fig:enlarged-H08}
\end{figure}

At H08, GCL$_E$ and GCL$_P$ remain visually similar over most of the domain in Figure~\ref{fig:enlarged-H08}. Physics-guided fine-tuning primarily produces localized modifications in precipitation intensity and spatial extent rather than a fundamentally different forecast field. This behavior is consistent with the weak and asymmetric construction of the physical regularization: the water budget information is intended to correct physically unsupported attenuation where necessary rather than replace the spatial structure learned by the data-driven model.

\subsection{Spatial-subset experiments}
\label{sec:subset-results}

The limited incremental improvement from physical regularization over the full enlarged domain motivates a more spatially resolved assessment. The spatial subset experiments examine whether the contribution of the moisture budget constraint becomes more visible when the forecasting problem is restricted to smaller spatial regions.

Table~\ref{tab:subset-results} reports the overall comparison between GCL$_E$ and GCL$_P$ for the spatial subset experiment.

\begin{table}[!htbp]
\centering
\caption{Overall forecasting performance in the spatial subset experiment.}
\label{tab:subset-results}
\resizebox{\textwidth}{!}{
\begin{tabular}{lrrrrrrrrrrrr}
\hline
Model &
MAE & RMSE & Bias & Corr &
CSI@4 & CSI@8 &
POD@4 & POD@8 &
FAR@4 & FAR@8 &
HSS@4 & HSS@8 \\
\hline

GCL$_E$ &
\textbf{0.2557} & \textbf{0.6582} & -0.0676 & \textbf{0.5508} &
0.0766 & 0.0708 &
0.0837 & 0.0765 &
\textbf{0.5241} & \textbf{0.5102} &
0.1401 & 0.1321 \\

GCL$_P$ &
0.2741 & 0.6669 & \textbf{-0.0139} & 0.5481 &
\textbf{0.0998} & \textbf{0.0919} &
\textbf{0.1195} & \textbf{0.1099} &
0.6235 & 0.6397 &
\textbf{0.1780} & \textbf{0.1680} \\
\hline
\end{tabular}
}
\end{table}

Restricting the spatial domain produces a substantially clearer physics-guided signal. Although GCL$_P$ does not improve MAE or RMSE, it reduces the magnitude of the negative bias from 0.0676 to 0.0139 mm h$^{-1}$, corresponding to a reduction of approximately 79\%. Event detection also improves consistently. CSI increases from 0.0766 to 0.0998 at 4 mm h$^{-1}$ and from 0.0708 to 0.0919 at 8 mm h$^{-1}$. Similarly, POD increases from 0.0837 to 0.1195 and from 0.0765 to 0.1099 at the two thresholds, while HSS increases from 0.1401 to 0.1780 and from 0.1321 to 0.1680, respectively.

These improvements are accompanied by higher FAR, showing that the physical constraint primarily increases sensitivity to precipitation events rather than uniformly reducing pointwise error. This behavior is consistent with its one-sided formulation, which penalizes physically unsupported attenuation but does not symmetrically suppress precipitation predictions.

\subsubsection{Robustness across spatial subsets and forecast horizons}
\label{sec:subset-robustness}

To determine whether the improvement observed for the central subset is specific to that region, the physics-guided constraint is further evaluated across the 16 spatial subsets defined above. Table~\ref{tab:subset-robustness} reports CSI and HSS during H05--H08 as mean $\pm$ standard deviation across the spatial regions.

\begin{table}[!htbp]
\centering
\caption{Moderate- and heavy-precipitation event skill at H05--H08 averaged over
16 spatial subsets. Values are reported as mean $\pm$ standard deviation across spatial
regions.}
\label{tab:subset-robustness}
\begin{tabular}{llcccc}
\hline
& & \multicolumn{2}{c}{CSI} & \multicolumn{2}{c}{HSS} \\
\cline{3-6}
Horizon & Model & @4 & @8 & @4 & @8 \\
\hline

H05 &
GCL$_E$ &
$0.1034\pm0.0609$ &
$0.0605\pm0.0511$ &
$0.1800\pm0.0975$ &
$0.1099\pm0.0879$ \\

&
GCL$_P$ &
$0.1093\pm0.0586$ &
$0.0589\pm0.0532$ &
$0.1897\pm0.0921$ &
$0.1066\pm0.0912$ \\[1mm]

H06 &
GCL$_E$ &
$0.0852\pm0.0568$ &
$0.0446\pm0.0454$ &
$0.1503\pm0.0922$ &
$0.0820\pm0.0791$ \\

&
GCL$_P$ &
$0.0935\pm0.0533$ &
$0.0452\pm0.0463$ &
$0.1643\pm0.0850$ &
$0.0827\pm0.0804$ \\[1mm]

H07 &
GCL$_E$ &
$0.0715\pm0.0544$ &
$0.0345\pm0.0428$ &
$0.1270\pm0.0891$ &
$0.0636\pm0.0750$ \\

&
GCL$_P$ &
$0.0811\pm0.0500$ &
$0.0373\pm0.0417$ &
$0.1437\pm0.0809$ &
$0.0689\pm0.0730$ \\[1mm]

H08 &
GCL$_E$ &
$0.0601\pm0.0525$ &
$0.0270\pm0.0401$ &
$0.1073\pm0.0869$ &
$0.0498\pm0.0711$ \\

&
GCL$_P$ &
$0.0707\pm0.0480$ &
$0.0317\pm0.0395$ &
$0.1260\pm0.0785$ &
$0.0586\pm0.0698$ \\
\hline
\end{tabular}
\end{table}

As expected, event-based skill decreases with forecast horizon for both models. At the 4 mm h$^{-1}$ threshold, however, GCL$_P$ achieves higher mean CSI and HSS than GCL$_E$ at every horizon from H05 through H08. At the higher 8 mm h$^{-1}$ threshold, GCL$_P$ is slightly lower at H05 but becomes higher from H06 onward, indicating that the contribution of physical regularization becomes more evident as the forecast horizon increases.

At H08, for example, mean CSI@4 increases from $0.0601\pm0.0525$ for GCL$_E$ to $0.0707\pm0.0480$ for GCL$_P$, while HSS@4 increases from $0.1073\pm0.0869$ to $0.1260\pm0.0785$. Improvements are also observed at the 8 mm h$^{-1}$ threshold, where CSI increases from $0.0270\pm0.0401$ to $0.0317\pm0.0395$ and HSS from $0.0498\pm0.0711$ to $0.0586\pm0.0698$.

Although the absolute differences are modest because moderate and heavy precipitation events become increasingly rare and difficult to predict at longer lead times, the behavior across the 16 spatial subsets shows that the physics-guided effect is not confined to a single selected region. Rather, its contribution is most consistently observed at later forecast horizons, where the underlying data-driven forecast progressively loses high-intensity precipitation signals.

\section{Discussion}

The results highlight complementary roles for the data-driven and physics-guided components of the proposed framework. While the geographically explicit architecture and occurrence-intensity-aware learning provide the most consistent improvements across experiments, the contribution of physical guidance is more selective and depends on forecast horizon and the spatial forecasting setting.

\subsection{Contribution of the data-driven architecture and event-aware learning}

Across the regional and enlarged experiments, the most stable improvement comes from the proposed spatiotemporal architecture and occurrence-intensity-aware learning. The basic GCL reduces systematic underestimation relative to GENESIS$_{\mathrm{FL}}$ in the regional experiment, while GCL$_E$ produces much larger gains in CSI and HSS at 4 and 8 mm h$^{-1}$. The enlarged domain 4-h comparison confirms that this advantage is not limited to the initial regional dataset: GCL$_E$ improves both pointwise accuracy and event detection despite the larger spatial domain and broader precipitation variability.

The regional component-wise comparison also provides evidence that the geographic representation and event-aware objective play distinct roles. The geographic representation improves the modeling of spatial variability, whereas the occurrence-intensity-aware objective primarily strengthens sensitivity to moderate and heavy precipitation events.

These gains come with a recurring FAR trade-off. GENESIS$_{\mathrm{FL}}$ often yields fewer false alarms at the higher thresholds, but this occurs together with much lower CSI and HSS. The pattern is consistent with conservative prediction of rare high-intensity events. In contrast, the occurrence-intensity-aware objective increases sensitivity to moderate and heavy precipitation, improving hits while also increasing false alarms. The categorical metrics should therefore be interpreted jointly rather than using FAR alone.

This distinction is particularly important for highly imbalanced precipitation data, for which low average prediction error or low FAR does not necessarily imply accurate representation of rare high-intensity events.

\subsection{Data-regime-dependent value of the physical constraint}

The incremental contribution of the moisture budget constraint is not uniform across experiments. On the full enlarged domain, GCL$_E$ and GCL$_P$ are nearly indistinguishable in most metrics; physics-guided fine-tuning mainly reduces negative bias. In the spatial subset experiment, however, GCL$_P$ produces consistent improvements in CSI, POD, and HSS, particularly during H05--H08, while also reducing the magnitude of negative bias.

This contrast is consistent with the intended role of the structured physical loss in Section~3.6. The physical term is weak and one-sided: it is designed to preserve physically supported precipitation when the data-driven forecast decays, rather than to replace the learned precipitation field. Its effect is therefore expected to be most visible when the baseline begins to lose moderate- and heavy-precipitation signals at longer lead times.

The reduction in negative bias observed after physics-guided fine-tuning is also consistent with this asymmetric formulation, since the constraint acts primarily against physically unsupported precipitation attenuation rather than symmetric pointwise forecast error.

The timing of constraint application also appears to depend on the data regime. For the full enlarged domain experiment, applying the physical constraint from the beginning or only at later lead times (H05--H08) produced similar results, with a slight advantage for the latter. In the spatial subset experiment, however, introducing the constraint too early occasionally destabilized training, as the physical term could exert excessive influence on the prediction. Applying the constraint at later lead times therefore provided a more stable setting while retaining its benefit for the stages at which precipitation skill deteriorated most clearly.

The subset experiment supports this interpretation, although the current experiments do not isolate spatial extent from all other aspects of data diversity and should not be interpreted as proving a universal dependence on domain size. The observed difference in training stability should therefore also be interpreted as a data-regime-specific behavior rather than as direct evidence that smaller domains necessarily require later physical guidance.

Taken together, these results suggest that both the effectiveness and the appropriate timing of physical guidance are data-regime dependent rather than uniform. It becomes more apparent when the data-driven forecast contains less information or progressively loses precipitation event skill at longer lead times.

\subsection{Implications and limitations}

Several limitations are important for interpreting the comparisons. First, GENESIS$_{\mathrm{FL}}$ is a reconstructed benchmark rather than an exact reproduction of the original published system. Second, the regional and enlarged experiments use different predictor configurations and forecasting settings, so absolute scores should not be compared directly across experiments; the primary evidence comes from within-experiment comparisons under matched settings. Third, the improvement in event detection from both event-aware and physics-guided learning is accompanied by increased FAR, indicating that future work should further refine the balance between missed heavy-rainfall events and false alarms.

In addition, the atmospheric water budget diagnostic is itself an approximate representation of precipitation-supporting conditions and may contain spatial, temporal, and physical-model uncertainties. The proposed one-sided regularization is designed to limit the influence of such uncertainty, but further evaluation across different climatic regions and precipitation regimes is needed to establish its broader robustness.

The relatively small differences between GCL$_E$ and GCL$_P$ over the full enlarged domain should also be interpreted cautiously; future work could quantify the uncertainty of these differences through repeated training runs or resampling-based statistical comparisons.

Overall, the experiments support two complementary conclusions. The geographically explicit CNN--ConvLSTM and occurrence-intensity-aware objective provide the most consistent improvement across datasets, while the moisture budget constraint acts as a targeted long-lead regularizer whose benefit is clearer when the data-driven forecast loses heavy-precipitation skill.

\section{Conclusion and Future Work}
\label{sec:conclusion}

This study developed a geographically explicit, occurrence-intensity-aware, and physics-guided spatiotemporal framework for hourly precipitation forecasting. The proposed approach addresses three complementary challenges: representing spatial heterogeneity beyond local convolutional dependence, learning effectively from the strongly imbalanced distribution of precipitation, and incorporating imperfect physical information without treating it as deterministic truth. Geographic information is introduced through a lightweight spatial embedding, while the occurrence-intensity-aware objective combines robust intensity regression, threshold event learning, and precipitation occurrence prediction. During the second-stage fine-tuning, atmospheric water budget information is incorporated through a weak asymmetric regularization strategy designed to discourage physically unsupported precipitation attenuation.

The experiments show that the geographically explicit architecture and occurrence-intensity-aware learning provide the most consistent improvements across the different forecasting settings. Explicit geographic representation improves the modeling of spatial variability, while event-aware learning substantially increases the Critical Success Index (CSI) and Heidke Skill Score (HSS) for moderate and heavy precipitation. Physics-guided fine-tuning has a more selective effect. Over the full enlarged domain, its primary contribution is a reduction in systematic negative bias, whereas over spatially restricted subsets it more consistently improves precipitation event prediction at later forecast horizons. These results suggest that physical guidance is most useful as a targeted, data-regime-dependent regularizer, with its contribution becoming more evident at later lead times when event skill deteriorates. They also highlight an important trade-off: increased sensitivity to moderate and heavy precipitation can be accompanied by higher false alarm rates.

Several directions remain for future work. The framework should be evaluated across additional climatic regions and precipitation regimes, while improving the trade-off between detecting high-intensity events and controlling false alarms. Further studies could isolate the contributions of the individual physical constraints and develop adaptive regularization strategies across forecast horizons. Statistical uncertainty could also be assessed through repeated training runs or bootstrap-based comparisons. More broadly, the asymmetric physics-guided formulation could be extended to other atmospheric variables and physical diagnostics.

\section*{Funding Information}
This work was supported by King Abdullah University of Science and Technology; the National Natural Science Foundation of China (No. 12271286); the Hebei Natural Science Foundation (No. A2025105010); and the Fundamental Research Funds for the Central Universities (No. XSQD-6120220282). 

\section*{Conflict of Interest}
The authors declare no conflicts of interest.

\section*{Data Availability Statement}
The ERA5 reanalysis data used in this study are publicly available from the Copernicus Climate Data Store at https://cds.climate.copernicus.eu. Guidance on accessing the ERA5 reanalysis data and the analysis code supporting the findings of this study are publicly available at https://github.com/xinyu-wang-0325/Spatiotemporal-Precipitation-Forecasting.

\end{document}